# Prediction of geophysical properties of rocks on rare well data and attributes of seismic waves by machine learning methods on the example of the Achimov formation

Dmitry Ivlev, *Zarubezhneft JSC*
dm.ivlev@gmail.com

**Abstract**

Purpose of this research is to forecast the development of sand bodies in productive sediments based on well log data and seismic attributes. The object of the study is the productive intervals of Achimov sedimentary complex in the part of oil field located in Western Siberia. The research shows a technological stack of machine learning algorithms, methods for enriching the source data with synthetic ones and algorithms for creating new features. The result was the model of regression relationship between the values of natural radioactivity of rocks and seismic wave field attributes with an acceptable prediction quality. Acceptable quality of the forecast is confirmed both by model cross validation, and by the data obtained following the results of new well.

**Introduction**

Anisotropy of properties of the reservoirs and seals for productive pools is one of the main uncertainties affecting the evaluation and further development of oil and gas production projects. This is especially true for the projects with hydrocarbon reservoirs in Achimov sediments at an early stage of exploration maturity, when there is on average little information and it consists of the test results from several wells and 3D seismic data. At this stage, seismic data is the main and almost the only source of knowledge about the area.

To predict the geological rock section based on seismic data, there have been developed various approaches to seismic-stratigraphic interpretation, which includes the following methods and approaches: amplitude analysis, geostatistical analysis, cluster analysis, seismic facies zonation, time thickness analysis, analysis of constant time slices, etc. Complexity and multifactorial nature of interpretation of the geological section by seismic data is noted in the papers [1, 2, 3, 4, etc.]. The complexity is associated with spatial resolution of the source seismic data, variability of the geological section and difficulty of transferring functions from one seismic survey to another even within the same field. Besides, a judgmental factor plays an important role in the interpretation of seismic data. All these factors create a combination of aleatoric and epistemic uncertainties in seismic-stratigraphic interpretation of a complex rock section, especially at an early stage of field maturity.

The work objective is to implement a comprehensive approach to reducing these uncertainties with subsequent prediction of spatial variability of geophysical properties of rocks by finding a relationship between various seismic attributes, their derivatives and geophysical parameters of the geological section defined in the drilled wells. Thereafter to reconstruct regression model between the sparse spatially localized high-resolution data and the encompassing hyperspace of the environment parameters with low spatial resolution using machine learning algorithms.

The object of the research is the productive intervals of Achimov sedimentary complex in the part of oil field located in Western Siberia. Papers of many authors are devoted to the conceptual model of the structure of Achimov sedimentary complex [5, 6, 7, etc.]. Deposition of Achimov sediments is associated with the ingress of silt and sandy sediments in the form of turbidite flows of different density and landslides to the footing of Neocomian shelf terraces. These phenomena are genetically and spatially connected with the areas of sediments discharge transported by fluvial distributary systems. Achimov thickness is considered in unity with its shelf zones as a single genetically connected complex and represents a fondoform part of the cycle. The cycle was developed in a relatively deep-water part of the basin, which is confirmed by many factors. Deposition of siltstone and sandstone layers was controlled by the activity of avalanche sedimentation zones - turbidite flows, landslides, and channels of transportation of sedimentary material that was coming from the continental part (canyons on the surface of the shelf terrace slope) [8, 9].

3D seismic data were acquired within the research area with subsequent processing and interpretation of acquired materials. Five wells were drilled in this area at different times: two exploration wells, two development horizontal wells and one development deviated well.

The following work sequence was implemented in the research: creation of the data set, creation of a regression model population, evaluation of attribute contribution to the forecast, combining the models into a metamodel, final reconstruction of the regression relationship and assessment of prediction quality.

**Creation of the basic data set for learning**

The main stage in creating a basic data set for learning (Fig.1) is the features generation [10]. Features generation is the process of using the domain knowledge to extract the features from the source data.

At the first stage, 25 algorithms of attribute extraction from seismic data were used for features generation; window sizes 5, 10, 20, 30, 50 were used for calculations. At the second stage, the gradient between the values obtained for each window and the frequency of spectral decomposition was calculated. Features with the best correlation with the determination factor 0.81 have been excluded from the resulting attribute data set for the studied area; a set of 341 attributes was obtained. At the third stage, four transform algorithms were applied to this data set. The first and the second derivatives were calculated, cubes clustering was carried out using k-mean algorithm, geomorphological attributes were calculated by upscaled cube slices (Fig. 1).

The project space was segmented by a 3D grid with 480x480x60 size, 5-meter vertical step and 25-meter horizontal step. The size was chosen as the minimum possible for accumulation of seismic data, as a smaller scale would not reflect the environment variability, assigning one value to a number of cells, and a larger size would lead to loss of information. Attribute values were normalized for the entire cube and brought to the grid size by the following methods: arithmetic mean, root mean square (RMS), maximum, minimum. After normalizing the values to the grid size, the features were filtered by the determination factor 0.81. A data set consisting of 4,084 attributes in each cell of the research area was acquired.

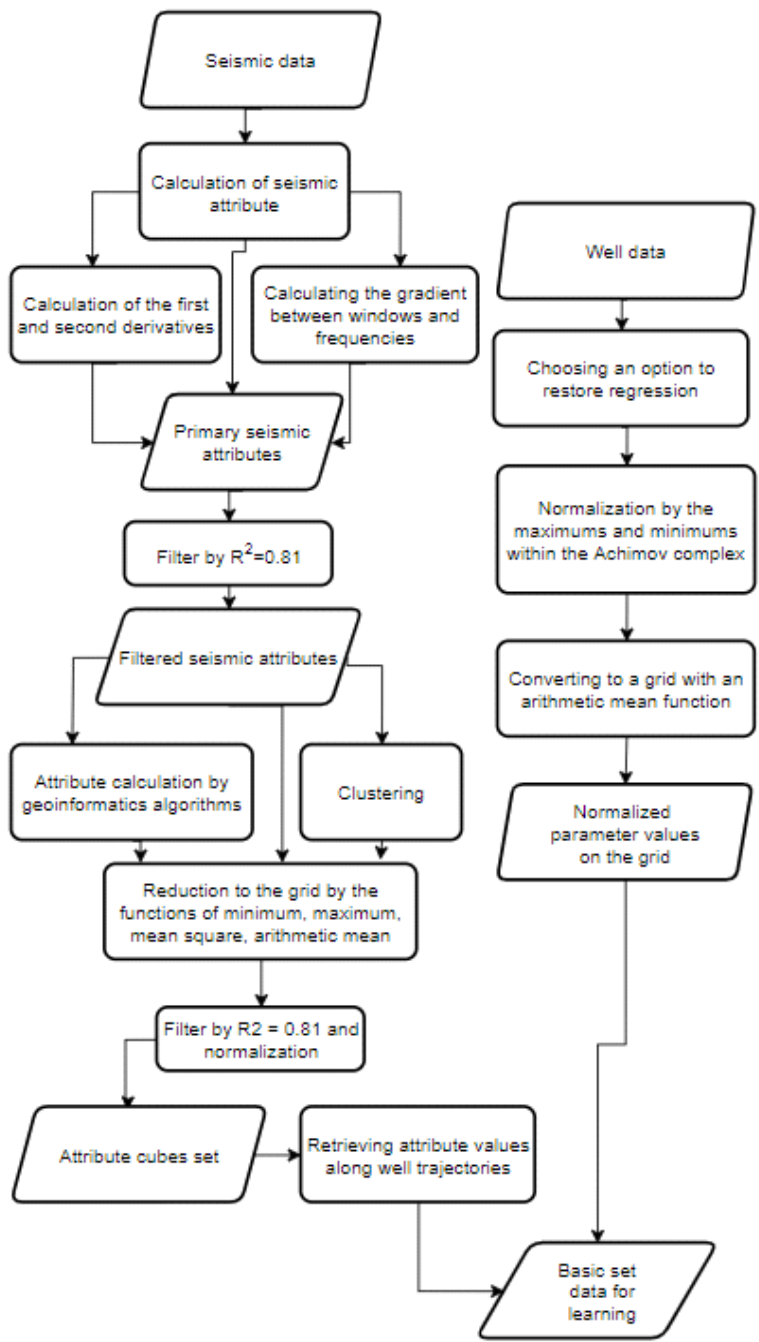

**Fig. 1.** Scheme for creating a basic training dataset

Selection of the petrophysical rock property as the target function of regression reconstruction is driven by the coverage with well logging methods in the drilled wells in the study interval and informative value of the method in terms of lithological interpretation of the geological section. Only one method fully satisfied the specified conditions - the method of measuring natural radioactivity of rocks (Gamma-Ray Logging) [11].

GR log was normalized to maximum and minimum measurements in the interval of Achimov complex and to its arithmetic mean value brought to the 3D grid. After normalizing, the target parameter was represented by a sample of 366 values averaged over a grid of normalized GR values.

**Generation of the synthetic data set – geophysical data augmentations**

Exploration analysis by various algorithms of machine learning showed unsatisfactory prediction quality and impossibility to make a robust generalization on the chosen wells with acceptable quality (Table 1). There are two possible options to enhance the sample by the target feature: drill additional wells and/or generate a synthetic data set (augmentations). The approach of generating a synthetic data set using two methods was used in this research paper (Fig. 2).

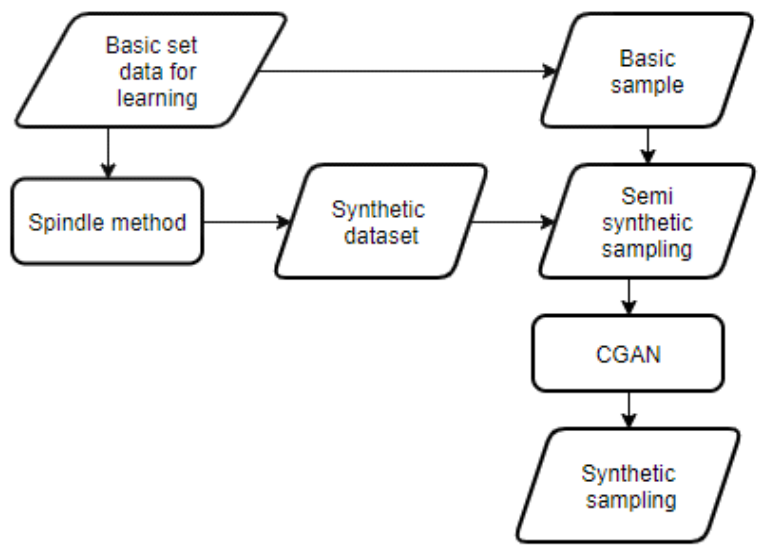

**Fig. 2.** Synthetic dataset creation scheme

The first method is the "spindle" method. The method got its name due to its implementation features: a well trajectory gets slightly deformed, and grid refinement to one meter is carried out locally near the well, average GR values and attribute values are calculated based on the log curve from the slightly changed new position, thus "winding on" the information surrounding the wellbore. The premise for this approach was the assumption on the uncertainty of the wellbore location in the wave field and the fact that various attribute parameters of the wave field reflect the pickups from different scale from lithology parameters detected by the GR curve. Thus, a data set of 1,697 GR values with 4,084 features was generated.

The second method is based on generation of the synthetic sample using the generative adversarial network based on the CGAN architecture (Conditional Generative Adversarial Nets).

Generative adversarial network (GAN) is the machine learning algorithm without a teacher, built on the combination of two neural networks, one of which (generative network) generates samples, and the second (discriminating network) tries to distinguish the right ("true") samples from the wrong ones. Generative and discriminating neural networks have the opposite goals - to create samples and to reject samples, an antagonistic game appears between them [12, 13].

Conditional Generative Adversarial Nets is a modified version of GAN algorithm, the architecture of which can be designed with help of transfer of additional data that is a condition both for the generator and for discriminator. Additional data can be transferred to the architecture in the form of any additional information, for example, a class mark, image or data from other models, which can allow to control the process of data generation [14].

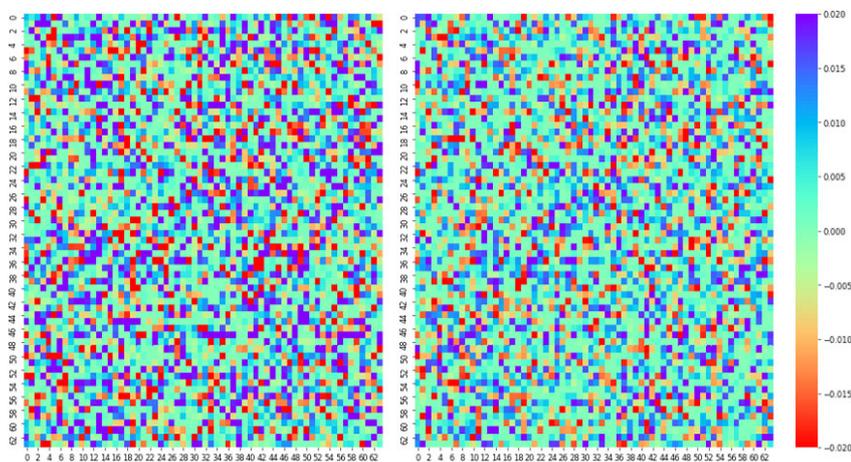

**Fig. 3.** Normalized attribute values reduced to a 2D set for CGAN learning.

Additional data in CGAN learning were the values of the target function - GR log. Initial vector of 4,084 combinations of normalized attributes reduced to the 2D set 64x64 was used as the learning data; the missing elements were duplicated (Fig. 3). Any number of synthetic data can potentially be generated using this method, as the quality of the resulting set of attributes for each GR value will be determined by its representation in the learning data set. After algorithm refining for this specific task and architecture learning, a data set was generated for 5,000 values of the target function in accordance with its initial distribution.

Three learning samples were formed following the results of synthetic data generation:
- basic - it includes just the initial data set with 366 GR values with 4,084 features;
- semi-synthetic - basic plus a set of data acquired using the spindle method 2063x4084;
- synthetic - semi-synthetic plus data generated by CGAN algorithm 7063x4084;

Total sampling is 7,063, which is equivalent to 118 vertical wells brought to the grid with vertical resolution 60 cells.

**Creation of regression models population and evaluation of attribute contribution to prediction**

Gradient boosting algorithm over decision trees in its specific implementations CatBoost [15], LightGBM [16] and XGBoost [17] was used to restore the regression relationship. Presently, this class of algorithms is a proven standard for learning classification and regression models on a heterogeneous set of numerical and categorical data. Use of third-party libraries SHAP [18] and LIME allows to interpret the resulting models.

Huber function was used as the loss function, which is a combination of mean squared error function (RMS) and mean absolute error function (MAE).

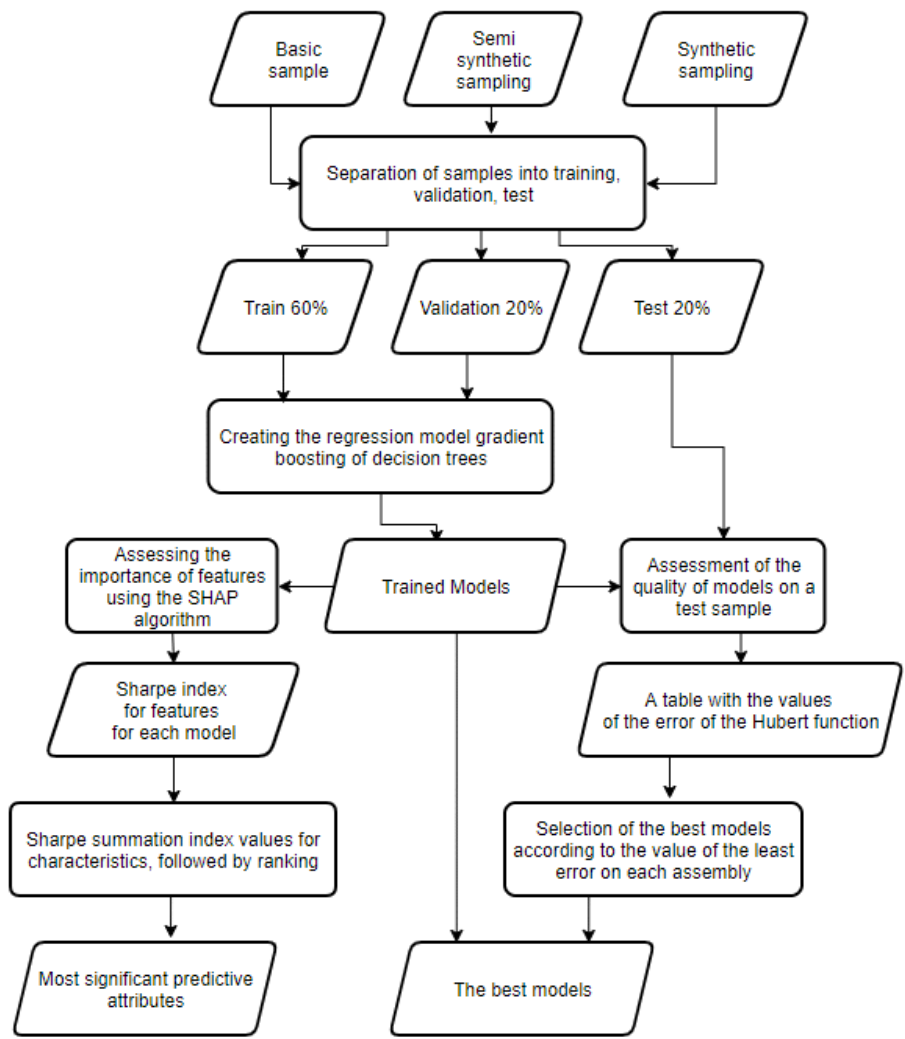

**Fig. 4.** Scheme of generating regression models population and evaluation of attribute contribution

Each sample was divided into three parts: training, test, and validation in the proportion of 60x20x20, monitoring that at least 40% of values from the basic or semi-synthetic samples are present in the testing or validating data set (Fig. 4).

Configuring hyper-parameters of machine learning algorithms was carried out on semi-synthetic sample and was used for all experiments.

**Table 1.** Hubert error function values

| Sample / Model | Catboost | LightGBM | XGBoost | Average |
|---|---|---|---|---|
| Basic | 0.1375 | 0.1281 | 0.1687 | 0.1448 |
| Semi-synthetic | 0.0588 | 0.0616 | 0.0583 | 0.0596 |
| Synthetic | 0.0159 | 0.0157 | 0.0151 | 0.0156 |

Table 1 shows the average result of obtained values of Huber error function on the test sample after 100 iterations of learning each algorithm by the samples with their various random separation.

Quality of the algorithm prediction on the test samples is on average the same for each of the three data sets.

Enriching the initial sample with synthetic data at each stage significantly reduced the error and its variance. Adding the data acquired by the spindle method to the basic sample reduced the error 2.43 times on average, and increase of the sample due to the data generated by CGAN additionally reduced the error 3.82 times. The achieved prediction quality is good, given the initial level of error on the basic sample.

Attribute contribution to prediction was evaluated for each of 900 models. Features relevance was evaluated by calculation of the Shapley index vector for them. Values of the Shapley index for each feature were summed and sorted in descending order of their impact on prediction.

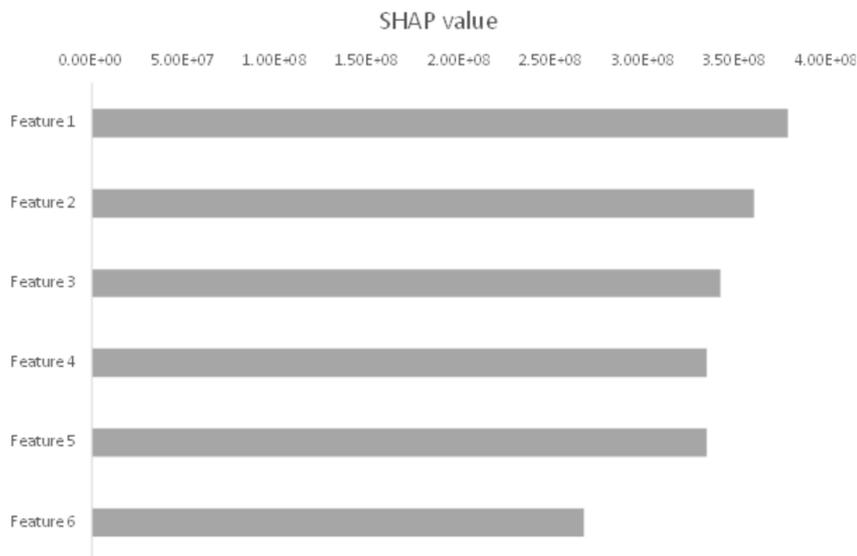

**Fig. 5.** Shapley value for the six most important attributes

The first six attributes most significant for prediction are shown here (Fig. 5). The others were inexpressive relative to each other in terms of impact degree; thus, the total Shapley index of the seventh feature was already 29% less than the sixth one, and the change in contribution of the subsequent features to the 542nd made an almost uniform contribution to the prediction as a steadily decreasing function with 0.3% slope; the impact of the remaining features was insignificant in the aggregate but was sometimes material for single models.

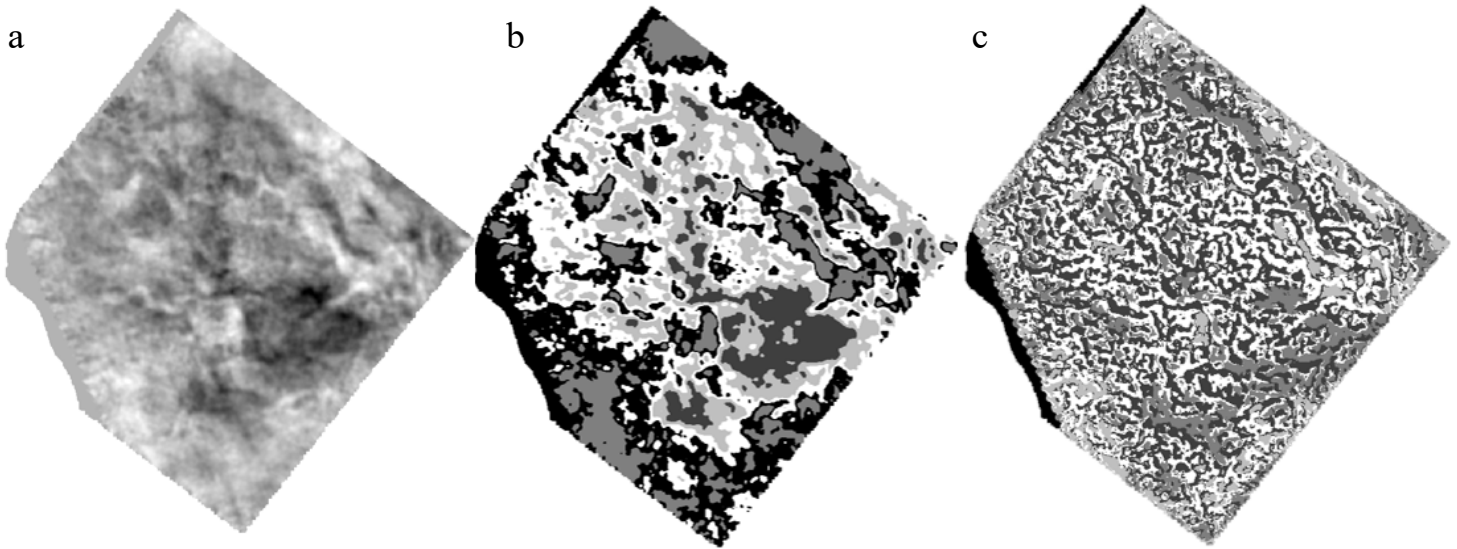

**Fig. 6.** Comparison of space clustering methods: a) horizontal slice according to the values of the spectral decomposition cube; b) clustering by the k-mean algorithm; c) clustering with the terrain clusters algorithm

Feature with ranking 1 was calculated by the terrain clusters algorithm [19] with 5 clusters extracted from the slices of spectral decomposition cube of the original seismic cube using Ricker wavelet with 30 Hz frequency, 1 second pulse length and a scale factor 4 reduced to the average value grid.

The idea of using this algorithm was born in process of searching for methods to formalize the "geological view" on the seismic attribute slices, when the geological bodies are picked by attribute slices, their gradient of changes in values and their shape (Fig. 6a). The k-mean clustering algorithm is not capable of such separation, as it works directly with attribute values [20]. At the same time, if we consider this slice as a proxy terrain - reflecting the geomorphology of the paleo-object, it is possible to divide this space using the methods of geoinformatics, as part of which there was developed a wide range of techniques for segmentation of landforms, one of which is Terrain Clusters. Fig. 6 shows the difference in splitting the space into 5 classes using two clustering algorithms, the original horizontal slice of spectral decomposition is shown on the left (Fig. 6a), the same slice after clustering by k-mean algorithm (Fig. 6b) is in the center, and the result of terrain clusters algorithm is on the right (Fig. 6c).

Feature 2 is the first derivative of sweetness seismic attribute in its arithmetic mean approximation on the grid.

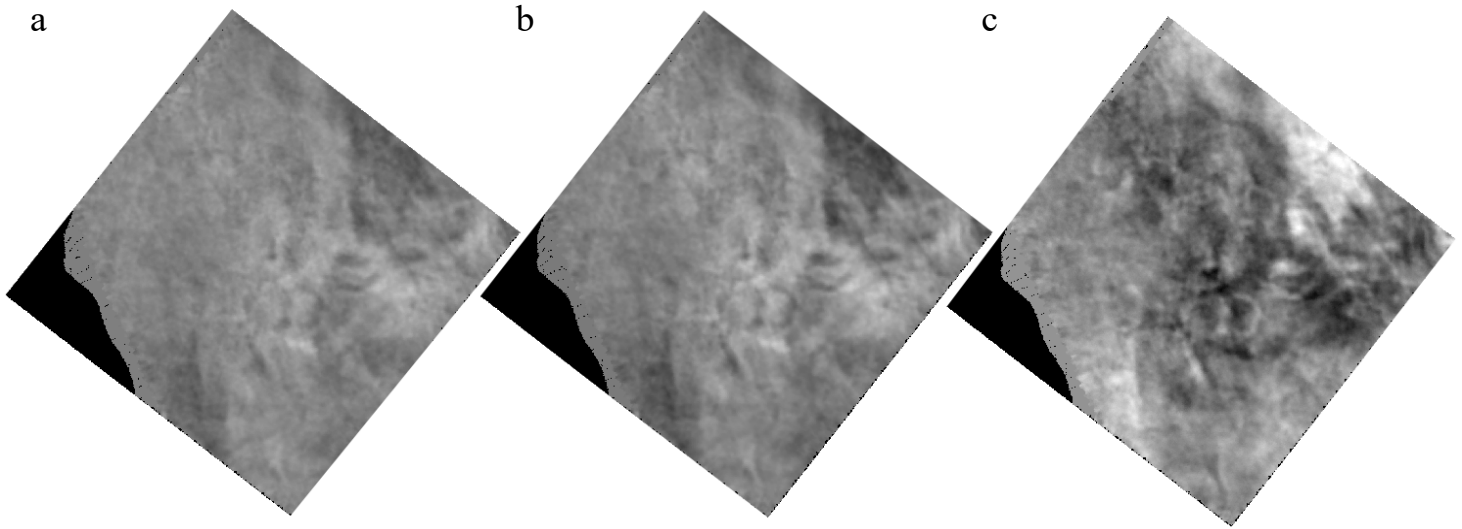

**Fig. 7.** The gradient between the frequencies of the spectral decomposition: a) frequency 20 Hz; b) frequency 5 Hz; c) the gradient between these frequencies

Feature 3 is the gradient between the frequencies from 20 to 5 Hz of the spectral decomposition by the Ricker wavelet with 1 second pulse length and a scale factor 4 with reduction to the grid by RMS value. Using a gradient allows to minimize the values of reiterated shapes and enhance the changes between the windows and attribute picking frequencies (Fig. 7).

Feature 4 is calculated by k-mean method with allocation of 21 clusters by amplitude spectrum attribute with grid reduction by maximum value.

Feature 5 is the tie to one of five stratigraphic horizons.

Feature 6 is the instantaneous frequency attribute, reduced to the grid by the average value.

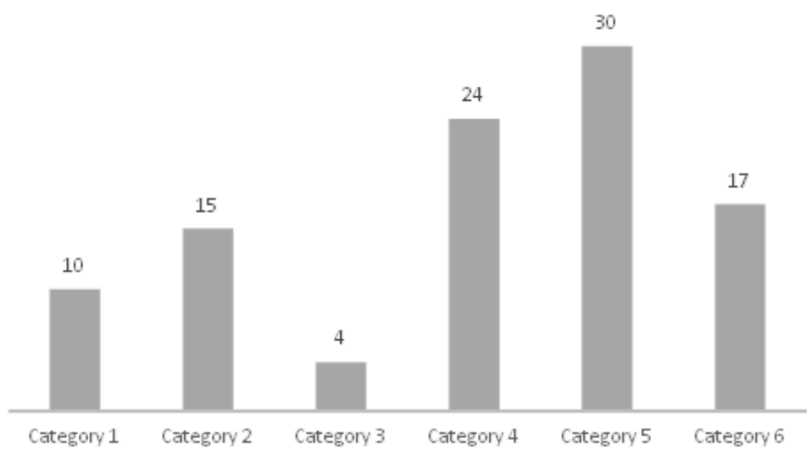

**Fig. 8.** Distribution of 542 Significant Attributes by Conversion Categories

The graph (Fig. 8) shows the percentage of the first 542 significant features distributed between the categories according to the transformation method. The first category is formed by the original attributes extracted from the seismic data and the expert estimates, which is 10% of 542. The second and the third categories are the first and second derivatives - 15% and 4% respectively. The fourth category includes the features created by the k-mean 24% clustering algorithm. The fifth includes features created by the methods of geoinformatics 30%. The sixth is represented by the method of calculating the gradient between attribute values extracted by different windows and frequencies of spectral decomposition.

**Ensembling the best models into the metamodel, the final forecast with quality assessment**

For the final forecast (Fig. 9), models with minimum values of the Hubert error function were combined into a metamodel with a stack generalization by the RidgeCV linear regression algorithm [21]. Stack generalization uses the prediction of each individual "weak" model and combining the strong predictive properties of each model makes a potentially better-quality prediction.

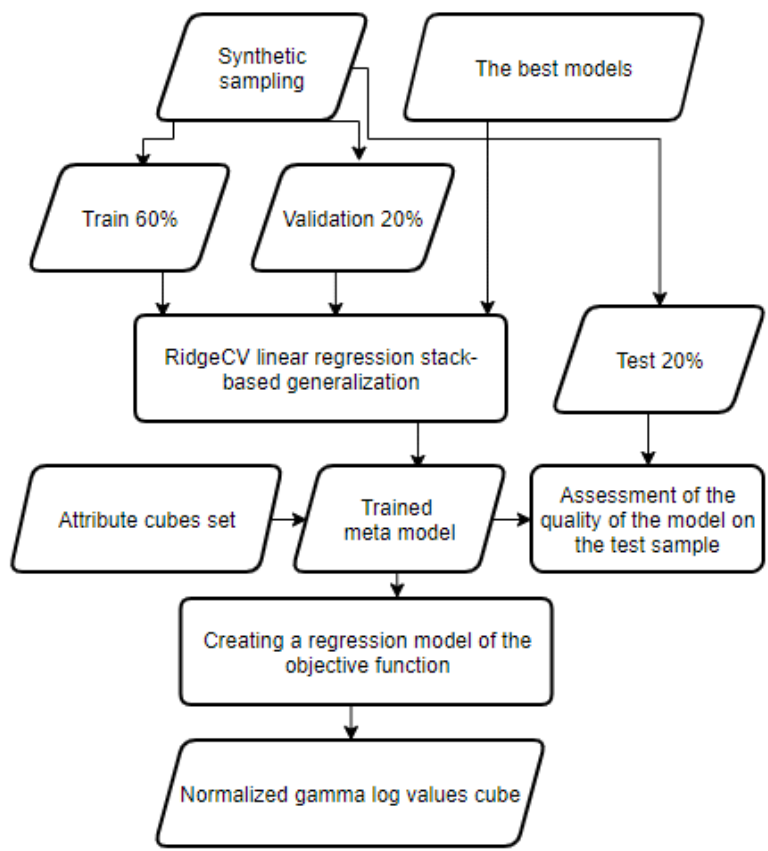

**Fig. 9.** Scheme of final prediction

Learning of the metamodel was carried out sequentially on three samples, first on the base case sample, then the model was trained on the semi-synthetic sample, and finally - on the synthetic sample. The final error of reconstructing regression model of the target function was 0.045. Relatively insignificant increment in the quality of metamodel prediction over individual models is probably due to the resolution limit of the original seismic data and their derivatives.

The metamodel performed prediction of normalized GR values over the entire volume of the survey space. Model prediction quality was checked using both actual drilling results of vintage wells and a new well data.

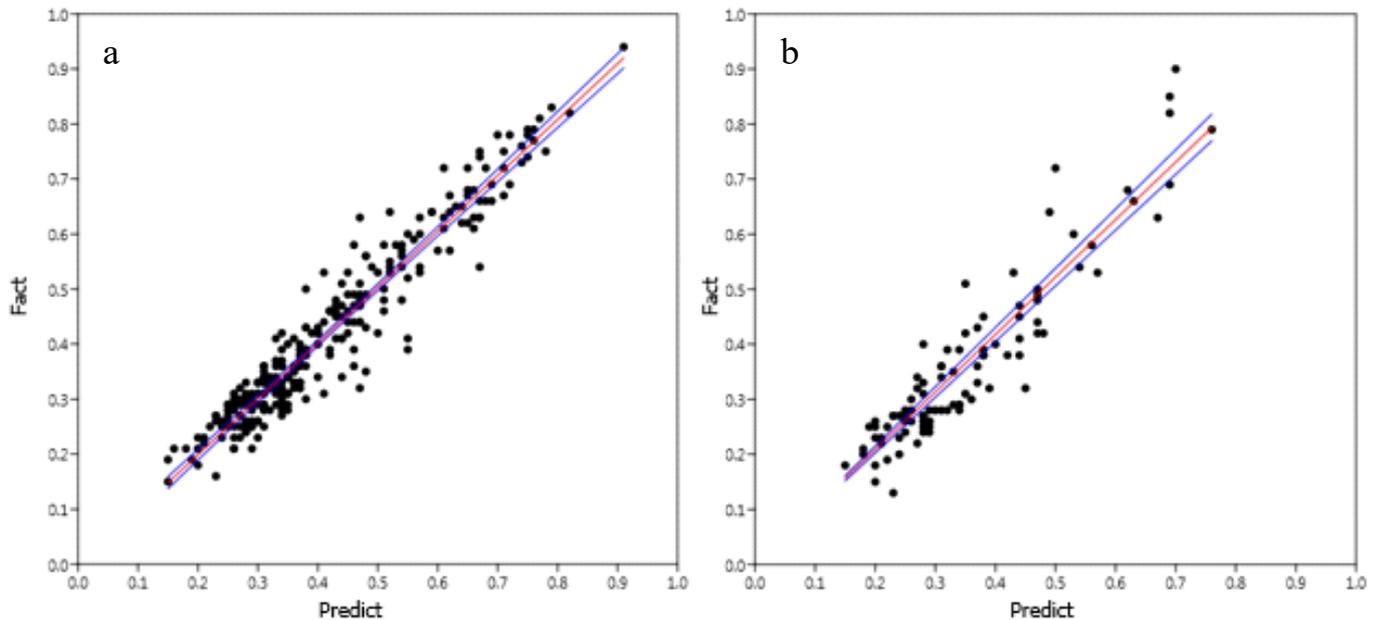

**Fig. 10.** The ratio of the predicted values of the normalized gamma-ray logging and the actual: a) wells from the training dataset; b) new well with horizontal completion

For the wells from learning sample (Fig. 10a), the correlation coefficient $R^2$ between predicted model values and actual values of normalized GR logs was $R^2 = 0.71$ with a standard deviation of 0.061. For a newly drilled well with a 1,000 m horizontal segment (Fig. 10b), 150 meters parallel from one of the wells from the learning sample, the data of which the model "did not see", the correlation coefficient dropped to $R^2 = 0.66$ with a standard deviation of 0.08. Despite the decrease in $R^2$, correlation between the absolute values of the forecast and actual data on the new well remained quite high; this indicates an acceptable predictive value of the model, at least in the area of 150 meters from wellbore locations from the training sample. It should be noted that it is likely that the correlation of absolute values between the fact and the forecast will decrease when moving away from these wells, this is due to different conditions of seismic acquisition and peculiarities of its processing, however, for the fields at an early stage of exploration maturity, in some cases, it is enough to keep the trend of changes in the forecast values, to assess the potential economic efficiency of field development and to adjust location of exploration wells and the wells for pilot production.

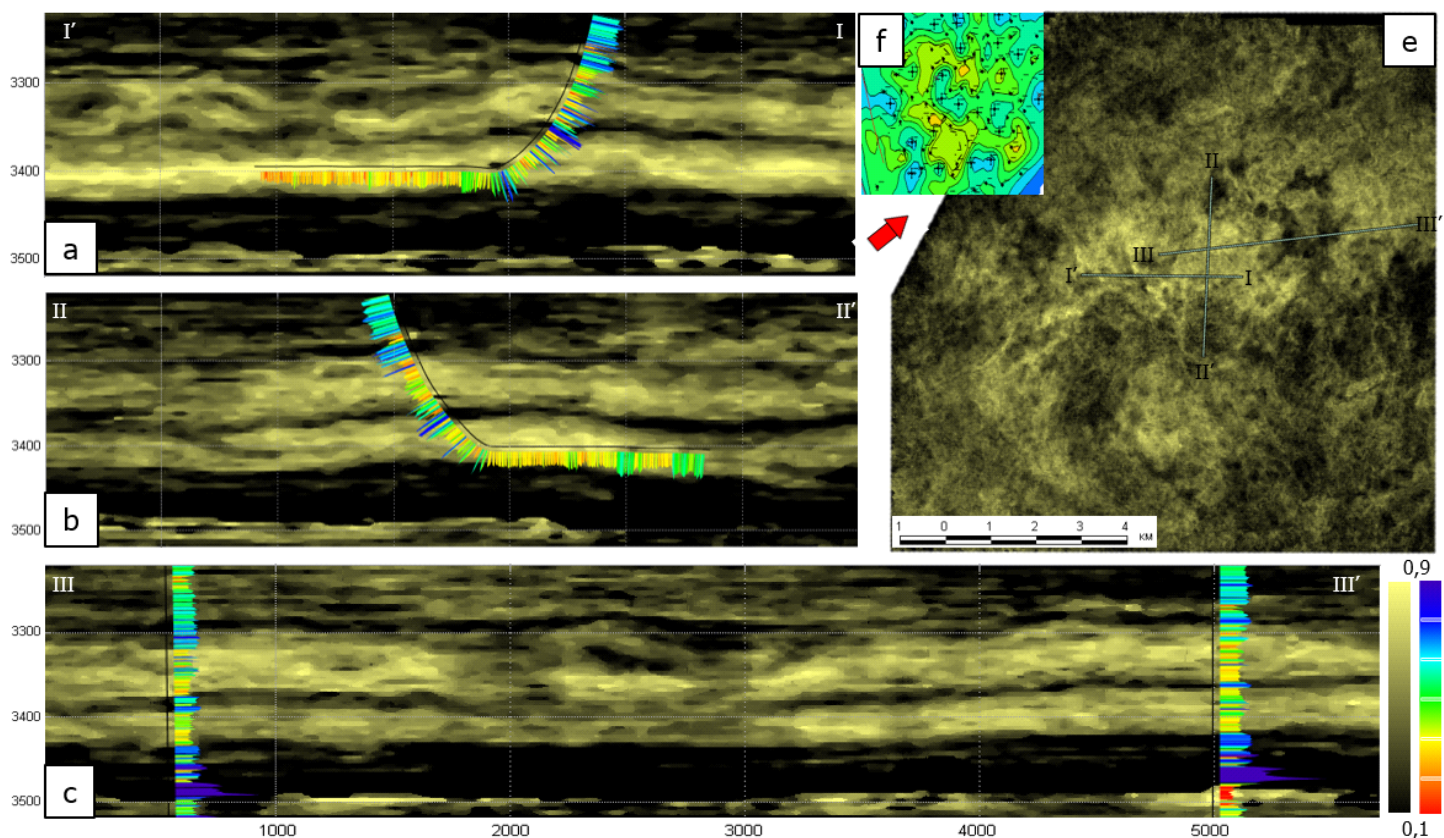

**Fig. 11.** Forecast of normalized rock properties: a), b), c) sections of porosity cube along the wellbore with GR log curve; e) map of total net pays for Achimov sedimentary complex; f) fragment of the map of total net pays according to the well data (data from wells that penetrated Achimov sequence – transit wells).

Regression relationship between GR values and effective porosity was restored by machine learning algorithms based on well data. Effective porosity cube was calculated.

Map of total normalized net pays (Fig. 11e) and sections of the projected cube of normalized porosity (Fig. 11a, 11b, 11c) demonstrate a complicated detailed picture of the Achimov complex structure. The sections show a good correlation between GR values in the wells and the projected properties cube; there is a complex sedimentological nature of argillaceous and sandy formations.

The structure of this sedimentary complex following the results of prediction is most likely typical. As an example, there is a fragment in one frame scale for the net pay map of Achimov horizon based on the transit wells of Verkhnekolik-Eganskoye field (Fig. 11f). The forecast net pay map shows the similarity of maximum and minimum distribution and the shape of geological bodies formed in similar depositional conditions.

**Summary**

- This paper presents a successful attempt to overcome the uncertainties in seismic-stratigraphic interpretation of the complex rock section with good accuracy for the early stage of field maturity.
- The deliverable included the model of restored regression relationship between the values of natural radioactivity of rocks and seismic wave field attributes with an acceptable prediction quality. Acceptable quality of the forecast is confirmed both by model validation with complete removal of some data from the learning process, and by the data obtained following the results of a new well drilled 150 meters away from the well from the learning sample. The regression relationship between the natural radioactivity of rocks and effective porosity of reservoirs was restored based on well tops data and log interpretation data - transition to reservoir properties of the target was carried out.
- The result was achieved with help of process stack consisting of machine learning methods, methods of enriching the source data with synthetic data, algorithms of creating new features using the function for regression model reconstruction as the target one, measurements of natural radioactivity of rocks, including for horizontal segments of wells.
- Two approaches were developed to enriching the source sample (geophysical data augmentations): spindle method and with help of Conditional Generative Adversarial Nets architecture (CGAN). The effect of reducing training error and improvement of model quality is noted when using the synthetic data set.
- Six features most significant for prediction were shown based on the calculation of Shapley index vectors for regression models. The top most relevant 542 attributes were distributed by transformation categories.
- The magnitude of slices generalization is shown by the methods of geoinformatics on pseudo-terrain built based on seismic attribute slices and their derivatives as a way to formalize the ideas on dynamic sedimentation processes.
- At the same time, we should note the importance of both attributes without transformation and those transformed by clustering methods using k-mean algorithm, with help of the first derivative and the gradient between calculation windows, and the gradient between frequencies for spectral decomposition. Stratigraphic breakdown of the section by intervals of Achimov horizon, Bazhenov and Tyumen formations is also important for prediction.
- Developing the proposed approach on the basis of already acquired results, it is possible to progressively restore fluid saturation and rock permeability regression models, and then using the created reservoir simulation model or machine learning methods to make a forecast of field development scenarios and their economic efficiency. Application and expansion of the proposed process stack opens up a wide potential for prediction of development of fields at an early stage of exploration maturity and for decision-making process. It becomes possible to create a real-time loop with positive feedback between the seismic data, drilling results, and economic strategy of the company.